\documentstyle[12pt]{article}
\newcommand{\bb}{\begin{equation}}
\newcommand{\ee}{\end{equation}}
\newcommand{\beqa}{\begin{eqnarray}}
\newcommand{\eeqa}{\end{eqnarray}}
\newcommand{\ra}{\rightarrow}
\newcommand{\dr}{\bigtriangleup\rho}

\newcommand{\as}{\alpha_{s}}
\newcommand{\ms}{\overline{MS}}

\newcommand{\zm}{X_{m}^{*}}
\newcommand{\zl}{X_{L}^{*}}
\newcommand{\zr}{X_{R}^{*}}
\newcommand{\zw}{X_{w}^{*}}
\newcommand{\zwms}{X_{w}^{MS}}
\newcommand{\zlms}{X_{L}^{MS}}
\newcommand{\zrms}{X_{R}^{MS}}
\newcommand{\mt}{m_{t}^{*}}
\newcommand{\lh}{\hat{\lambda}}
\newcommand{\mtb}{m_{t}^{B}}
\newcommand{\at}{\alpha_{t}^{*}}
\newcommand{\atb}{\alpha_{t}^{B}}
\newcommand{\atms}{\alpha_{t}^{MS}}
\title{\begin{flushright}
{\small PITT-TH-94-04 \\
June 1994\\[1cm]}
\end{flushright}
{\bf ON THE SCHEME DEPENDENCE OF \\
THE ELECTROWEAK RADIATIVE CORRECTIONS}}
\author{{\bf A. I. Bochkarev}$^{\dagger}$ $\;$ and $\;$ {\bf R. S. Willey} \\
  {\it Department of Physics and Astronomy} \\
  {\it University of Pittsburgh} \\
  {\it Pittsburgh, PA 15260}}

\date{}

\begin{document}

\maketitle
\vspace{0.3in}
\begin{center}
Abstract\\
\end{center}

We study the scheme dependence of the two-loop expression for the
$\rho$-parameter of the Standard Model in the heavy $t$-quark mass
limit $m_{t} \gg m_{w}$. In the $\ms$-scheme the two-loop
electroweak correction to the ${\cal O}(m_{t}^{2})$ term in $\dr$ is
found greater than the QCD $\as$-correction.

\vspace{1.5in}

\noindent $^{\dagger}$ On leave from: {\it Institute for Nuclear
Research, Russian Academy of Sciences, Moscow 117312, Russia}

\vfill \eject

To confront current and forthcoming precision electroweak (EW) data,
one-loop, and recently, two-loop contributions to various EW parameters
have been calculated. Accurate information about the EW radiative corrections
is required to determine the unknown parameters of the EW Lagrangian or
obtain constraints on possible extentions of the Standard Model (SM).
When evaluating higher-order radiative corrections it is important to know
the optimal parameter of the perturbative expansion, so that the coefficients
in power series of that parameter are small. To achive this one naturally
looks for the appropriate renormalization scheme. The $\ms$-scheme
may be advantageous for the higher-order corrections to the off-shell
quantities,
since one can solve renormalization group equations with non-zero
masses in that scheme \cite{msew}. In this note we consider the
$\rho$-parameter of the SM. Since the $t$-quark is reportedly
\cite{top} twice as heavy as the $W$-boson the perturbative series
for $\rho$ is predominantly an expansion in powers of $\alpha_{t}$
($\alpha_{t}\,\equiv\,g_{t}^{2}/4\pi$ with $g_{t}$ being the
${\em top}$ Yukawa coupling constant) and QCD coupling constant
$\as$. For the currently discussed value $m_{t} \simeq 174 GeV$ the
two coupling constants are rather close: $\alpha_{t} \simeq .08$ and
$\as(m_{t}) \simeq 0.11$. Below we convert the known two-loop
expression for $\rho$ into the $\ms$-scheme and find that in terms of
the running coupling constants $\alpha_{t} (m_{t})$ and $\as(m_{t})$
the EW ${\cal O}(\alpha_{t}^{2})$ corrections are greater than
the QCD ${\cal O}(\alpha_{t} \as)$ ones. We also obtain a relation
between the physical and $\ms$- scalar selfcoupling constant $\lambda$
and present the $\ms$-expression for the recently calculated
${\cal O}(\lambda^{2})$-correction to the decay of Higgs boson into
fermion-antifermion pair.

The $\rho$-parameter is the ratio of the amplitudes of neutral and
charged weak currents at low energies. In the SM $\rho$ differes
from its tree level value $\rho =1$ due to non-zero mass splitting
within the fermionic doublets \cite{vel}. The largest correction
comes from the $t$-quark. For $m_{t}\gg m_{w}$, $\rho$ is given by
the ratio of the $W$- and  $Z$-boson propagators at zero momentum:
\bb
\rho\;\;=\;\;\frac{1+\Pi_{w}(k^{2}=0)/m_{w}^{2}}{1+\Pi_{z}(k^{2}=0)
/ m_{z}^{2}} \label{rhodef}
\ee
where $\Pi_{w,z}(k^{2})$ are the transverse parts of the polarization
operators of $W$- and $Z$-bosons.
For $m_{b} \ll m_{t}$ one obtains for $\dr \equiv 1 - 1/\rho$ :
\bb
\dr\;\;=\;\;N_{c}\,\frac{\at}{8 \pi} \;\left[ 1\;-\; \frac{2}{9}\,
\frac{\as}{\pi}\,(\pi^{2}+3) \;+\;\frac{\at}
{8 \pi}\;\rho^{(2)} \right]
\label{dromsh}
\ee
where $N_{c}=3$ is the number of colors.
The QCD-correction ${\cal O}(\alpha_{t} \as)$ comes from \cite{qcd}.
Possible form of the futher QCD-corrections were discussed recently
in \cite{volsir}.
The EW correction ${\cal O}(\alpha_{t}^{2})$ is determined by
$\rho^{(2)}$, which is a function of the ratio $r = m_{H}/m_{t}$ of
the Higgs boson mass to the $t$-quark mass. It was computed
numerically in \cite{it} and analytically in \cite{tar}.
The expression (\ref{dromsh}) is
obtained in the on-shell momentum subtraction scheme (marked by the
superscript "*") \footnote{We will not be sensitive to the scheme
dependence of $\as$}. The parameter $\at$ is expressed in terms of
the measured quantities
\bb
\at\;\;=\;\;\frac{G_{\mu}^{*}}{\sqrt{2}}\,\frac{(\mt)^2}{\pi},
  \label{altmsh}
\ee
where the Fermi-decay coupling constant
$G_{\mu}^{*} = 1.16639(2) \cdot 10^{-5} GeV^{-2}$ \cite{gmu} is
defined
in such a way that pure electromagnetic corrections do not contribute
to it. As far as the virtual heavy $top$ and Higgs boson
contributions are concerned,
$G_{\mu}$ is determined by the $W$-boson polarization operator:
\bb
G_{\mu}\;\;=\;\;\frac{1}{\sqrt{2}\,\upsilon^{2}\,
(1+\Pi_{w}(k^{2}=0)/m_{w}^{2})}       \label{Gmu}
\ee
$\upsilon$ being the v.e.v. of the Higgs doublet:
$\upsilon\, =\, 2 m_{w}/g_{w}$.

$\dr$ is a physical quantity. It is made finite by the renormalization
of the Lagrangian parameters - masses and coupling constants. The
quark mass in (\ref{altmsh}) depends on the scheme in which both
EW and QCD calculations were performed, while the $G_{\mu}$
according to (\ref{Gmu}) depends on the scheme of the EW
calculations.
Specifying the scheme one chooses a relation between the bare
Lagrangian parameters (marked by the superscript "B") and the
renormalised ones.
For instance, the bare quark propagator $S(k)$ parametrized near
the pole as
\bb
S^{-1}(\hat{k} \ra \mt)\;\;=\;\;\zl \hat{k} P_{L}\;+\;\zr \hat{k}
P_{R}\;-\;\mtb
\zm
\label{prop}
\ee
with $P_{L,R} = (1 \pm \gamma_{5})/2$, determines the physical mass
in terms of the bare one: $\mt\;=\;\mtb\,\zm / \sqrt{\zl \zr}$.
Changing the on-shell
subtracted $X^{*}$'s for the $X^{MS}$'s one obtains a relation
between the pole-quark mass $\mt$ and $\ms$-mass $m_{t}$. The QCD-part
of the function $\mt [m_{t}]$ was computed on the one-loop level in
\cite{nar} and on the two-loop level in \cite{2lms}. The one-loop
relation
\bb
\mt \;\;=\;\;m_{t}\;\left(1 \;+\; \frac{4}{3}\,\frac{\as}{\pi}\; +\;
\frac{\as}{\pi}\, \ln(\mu^{2}/m_{t}^{2}) \right)   \label{mtqcd}
\ee
with the usual choice of the normalization point $\mu = m_{t}$ leads
to the following form of the ${\cal O}(\alpha_{t} \as)$ - correction:
\bb
\dr\;\;=\;\;N_{c}\,\frac{\at}{8 \pi} \;\left[ 1\;-\; \frac{2}{9}\,
\frac{\as}{\pi}\,(\pi^{2}-9) \;+\;\frac{\at}
{8 \pi}\;\rho^{(2)} \right]  \label{drmsqcd}
\ee

The $\at$ in (\ref{drmsqcd}) reminds us that the EW
${\cal O}(\alpha_{t}^{2})$ -correction is still given in terms
the on-shell quantities according to (\ref{altmsh}). The rather small
coefficient of the ${\cal O}(\as)$-term in  (\ref{drmsqcd}) and
possibly \cite{top}
close numerical values of $\as$ and $\alpha_{t}$
indicate that the ${\cal O}(\as \alpha_{t})$- and
${\cal O}(\alpha_{t}^{2})$-corrections are of the same order of
magnitude. To clarify this point we now obtain a complete
$\ms$-expression for $\dr$ in the SM.

The two-loop EW calculations of $\dr$ are done in terms of the bare coupling
constant $\atb$ and then, in the course of renormalization, are usually
expressed in terms of the on-shell renormalized Yukawa
coupling constant $\at$:
\bb
\at\;\;\equiv\;\;\atb \;  (\zm)^{2}\,/ \zw\,\zl\,\zr   \label{atb*}
\ee
To obtain (\ref{atb*}) one uses (\ref{Gmu}) with
$\zw \,\equiv \,1+\Pi_{w}^{B}(k^{2}=0)/m_{w}^{2}$ and
$m_{t}^{B} \,\equiv \,g_{t}^{B} \upsilon^{B}/\sqrt{2}$.
$X_{w}$ is the renormalization constant of the $W$-boson mass:
$m_{w}^{*,MS}\,=\,m_{w}^{B}\, \sqrt{X_{w}^{*,MS}}$ in the limit $m_{t} \gg
m_{w}$ \cite{hol}. Due to the following Ward identity \cite{it}
it is related to the self-energy of the
corresponding charged Goldstone bosons $\Pi_{\phi}(k^{2})$ :
\bb
\Pi_{w}(k^{2}=0)/m_{w}^{2}\;\;=\;\;\lim_{k^{2} \ra 0} \; \Pi_{\phi}(k^{2})/
k^{2}         \label{ward}
\ee
In the $\ms$-scheme meanwhile one has:
\bb
\atms\;\;\equiv\;\;\atb \,/\, \zwms\,\zlms\,\zrms  \label{atbms}
\ee
Evaluating the one-loop diagrams of self-energies $S^{-1}(k)$ of the
$t$-quark and the charged Goldstone boson $\Pi_{\phi}(k^{2})$
we obtain the factors $X^{*,MS}$, which determine the
corresponding schemes. For the ratio of the physical $\at$ and
$\ms$-parameter $\atms$ one has:
\bb
\frac{\at}{\atms}\;\;=\;\; \frac{\zwms\, \zlms\, \zrms}
{\zw\, \zl\, \zr} \;  (\zm)^{2} \label{*ms}
\ee
The $\ms$-bare renormalization constants have generic form
$X^{MS}\, =\, 1 + {\cal O}(1/\epsilon - \gamma_{E} + \ln (4\pi))$,
where $\epsilon$  and $\gamma_{E}$ are the conventional parameters
of the dimensional regularization. The factors $X^{*}$
 have nontrivial finite parts and a dependence on the
normalization point $\mu^{2}$.
Note that $X_{m}^{MS} = 1$, since there is no unltraviolet-infinite
renormalization of the mass counterterm at least on the one-loop
level.

The calculation of the
${\cal O}(\alpha_{t})$-corrections is done with the conventional EW
Lagranagian in a renormalizable gauge
in the gaugeless (gauge-invariant) limit $\alpha_{w} = 0$
\cite{it}, \cite{tar}. The corresponding theory is a
linear $\sigma$-model of a scalar doublet $\Phi$ interacting with a doublet of
fermions $(t,b)$, and only the $t$-quark is relevant for
$m_{t} \gg m_{b}$. For the EW relation between the pole
mass $\mt$ and the $\ms$-mass $m_{t}$ we find:
\bb
\mt \;\;=\;\; m_{t}\; \left( 1\;+\;\frac{\atms}{8\pi} \;
\left[\bigtriangleup m(r)
\,-\,\frac{3}{2}\,\ln(\mu^{2}/m_{t}^{2})\right]\right)     \label{mtmt*}
\ee
with
\bb
 \bigtriangleup m(r)\;\;=\;\;\int_{0}^{1}dx \, (2-x) \, \ln\left(r^{2}(1-x) +
 x^2 \right) \;- \;\frac{1}{2}     \label{mtew}
\ee
In contrast to QCD, the very notion of the $\ms$-mass $m_{t}$ in the
EW theory is not
determined entirely by the prescriptions of the minimal
subtraction scheme. It depends on the value of $\upsilon$ chosen
as a parameter of the calculations: $m_{t} = g_{t} \upsilon$.
Eqn. (\ref{mtmt*}) corresponds to the $\upsilon$, incorporating
the tadpole contributions (Fig. 1), which are nonzero in the
$\ms$-scheme (the details are in \cite{bw}):
\bb
\upsilon^{2}\;\;=\;\;\upsilon_{MS}^{2}\;\left(1\;+\;6 \,\lh \;-\;
\frac{2 N_{c}}{\lh}\,\left(\frac{\alpha_{t}}{8 \pi}\right)^{2}\right)
\label{v}
\ee
where $\lh\,=\,\lambda / (16 \pi^{2})$, $\lambda$ being a scalar
self-coupling constant and $\upsilon_{MS}$ - parameter of the
$\ms$-calculations.
For the relation (\ref{*ms}) between the on-shell- and the
$\ms$-Yukawa coupling constants one obtains (see Fig. 2):
\bb
\at\;\;=\;\;\atms\;\left[ 1\;+\;\frac{\atms}{8\pi}\;\left\{
2 \bigtriangleup m(r)\;-\;N_{c}\;-\;\frac{1}{2} r^{2}\;-\;
(3 + 2N_{c})\; \ln(\mu^{2}/m_{t}^{2})\right\} \right]  \label{a*ms}
\ee

The relation (\ref{a*ms}) between dimensionless quantities does not
require a calculation of the tadpole diagrams, corresponding to
the shift (\ref{v}) in $\upsilon$.
The term $\sim \; r^{2} \atms$ in (\ref{a*ms}) is essentially the
scalar self-coupling constant $\lambda$ originating from the
renormalization of $G_{\mu}$ according to eq.(\ref{Gmu}).
Thus we find a
${\cal O}(G_{\mu}^{2} m_{t}^{2} m_{H}^{2})$-correction in the total
$\ms$-expression for $\dr$, which reads as
\bb
\dr^{MS}\;\;=\;\;N_{c}\,\frac{\alpha_{t}}{8\pi}\;\left[1\;-\;
\frac{2}{9} \,\frac{\as}{\pi}\,(\pi^{2}-9) \;+\;\frac{\alpha_{t}}
{8 \pi}\;\left(\rho^{(2)}\;+\;2 \bigtriangleup m(r)\;-\;N_{c}\;-\;\frac{1}{2}
r^{2}\right) \right]  \label{droms}
\ee
where both $\alpha_{t}$ and $\as$ are $\ms$-parameters normalized at
$\mu^{2}=m_{t}^{2}$.  Because of the smallness of the coefficient
in the ${\cal O}(\as \alpha_{t})$-term the correction is dominated by
the ${\cal O}(G_{\mu} m_{t}^{2} m_{H}^{2})$-term ($r^2/2$-term in
(\ref{a*ms})) which is not present in the on-shell renormalized
$\dr^{MOS}$.

To fix the electroweak contribution, we take for illustration,
$m_t= 174 GeV$ (then $\at = .076$ and $\as(m_t)= .11$ ) and shift
$\at \ra \alpha_{t}$ for $m_{H} = 300 GeV$ Higgs boson.
The results are presented in two graphs, Fig. 3 and Fig. 4.
{}From these figures, one can see that both the EW term and
the QCD term have the same sign (negative) and the EW contribution
is substantially larger than the QCD
contribution, by a factor of $6$ for light Higgs to a factor of $15$ for a
Higgs mass of slightly more than $1\, TeV\;\; (r=6)$.
For $1 TeV$ Higgs, the two-loop EW correction is dominated by the ${\cal
O}(G_{\mu}m_t^2m_H^2)$-term ($\frac{1}{2}r^2$-term in eq.(\ref{droms})).
In the approximation we considered ($m_{t},m_{H} >> m_{w}$), $\dr$ as
defined in (\ref{rhodef}) is a physical quantity. Thus, if calculated
to all orders, its value should be independent of the
reparametrization involved in the change of the schemes. In finite
order there are unequal truncation errors, but these should be small
if perturbation theory is good in both schemes. We fix $\alpha_{t}$
by inverting (\ref{a*ms}) at some conventionally chosen value of $r$.
Since many SM loop correction calculations are done with an
illustrative value $m_{H} = 300 GeV$, we make the conversion at
$r=300/174$. Then we obtain a nontrivial $r$-dependent scheme
dependence (Fig. 4). Again one can see the domination for large
$r$ of the $r^2/2$-term in (\ref{droms}), introduced by the
transformation (\ref{a*ms}).

Consider now radiative corrections to the heavy Higgs boson decay into
fermion-antifermion pair \cite{marc}. The two-loop corrections
${\cal O}(\lambda^{2})$ have been computed recently with two
different answers \cite{wisc} and \cite{ghinc}. The later calculation
\cite{ghinc} gives the following correction $\bigtriangleup \Gamma$
to the decay rate $\Gamma (H \ra f \bar{f})$, performed within the
momentum-subtracted on-shell scheme:
\bb
\bigtriangleup \Gamma\;\;=\;\;1\;+\;2.1172\,\lh^{*}\;-\;19.4483\,
(\lh^{*})^{2}              \label{dg}
\ee
where the physical scalar coupling
constant
$\lambda^{*}\;\equiv\;G_{\mu}^{*} (m_{H}^{*})^{2} / \sqrt{2}$. To
get the $\ms$-version of (\ref{dg}) we have calculated a
relation $\lambda^{*} [\lambda^{MS}]$. Following the same strategy as
used above for $\at [\atms]$ we obtain on the one-loop level
(see details in \cite{bw}):
\bb
\lh^{*}\;\;=\;\;\lh^{MS} \;\left( 1 \; -\; \left[ 25\,-\,3\pi\sqrt{3}
\;+\; 12 \ln(\mu^{2}/m_{H}^{2})\right]
\; \lh^{MS} \right)            \label{lms}
\ee
Hence the $\ms$-expression for $\bigtriangleup \Gamma$ numerically
reads as
\bb
\bigtriangleup \Gamma\;\;=\;\;1\;+\;2.1172\,
\lh^{MS}\;-\;37.8167\,(\lh^{MS})^{2}  \label{dgms}
\ee
for $\mu^{2} = m_{H}^{2}$. The coefficient of $\lh^2$ appears to be
larger in $\ms$ than in $MOS$-scheme.

In conclusion, we have found that while in the on-shell subtraction
scheme the QCD ${\cal O}(\alpha_{t} \as)$-correction dominates
and there is sensitivity to the heavy top quark mass, in the
$\ms$-scheme the EW ${\cal O}(\alpha_{t}^{2})$-correction dominates
and there is sensitivity to the heavy Higgs mass.

\section*{Acknowledgments}

We would like to thank A. Duncan for discussions.
This work was supported by the U.S. National
Science Foundation under grant and PHY90-24764.

\newpage
\section*{Figure Captions}

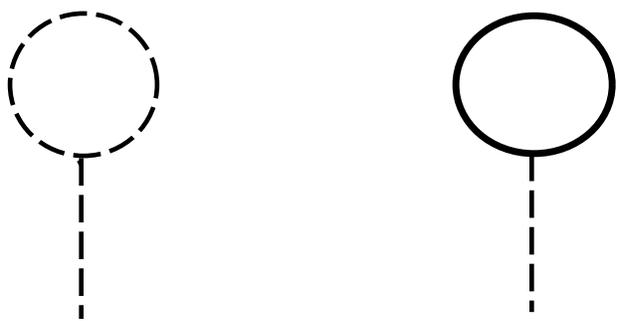
\noindent Fig. 1. Tad-pole diagrams responsible for the shift of the
vacuum expectation value $\upsilon$ in $\ms$-scheme from its
tree-level value: dashed line - Higgs boson loop, solid line -
$t$-quark loop.\\
\noindent Fig. 2. The ratio $\mt / m_{t}^{MS}$ (solid line) and the
ratio $\at / \atms$ (dashed line) as functions of $r$ in the interval
of $r$ corresponding to the Higgs masses
$m_{H} \simeq \{60 GeV, 1 TeV\}$.\\
\noindent Fig. 3. QCD (dashed line) and EW (solid line)
corrections to the one-loop expression for $\dr^{(1)}$ as a percent
of $\dr^{(1)}$ in the $\ms$-scheme. \\
\noindent Fig. 4. $\dr$ as a function of $r$ on the
$2$-loop level in the $MOS$-scheme (dashed line) and
$\ms$-scheme (solid line). Dotted line correspondes to using
$\ms$-scheme in QCD calculations only (eq. (\ref{drmsqcd}).

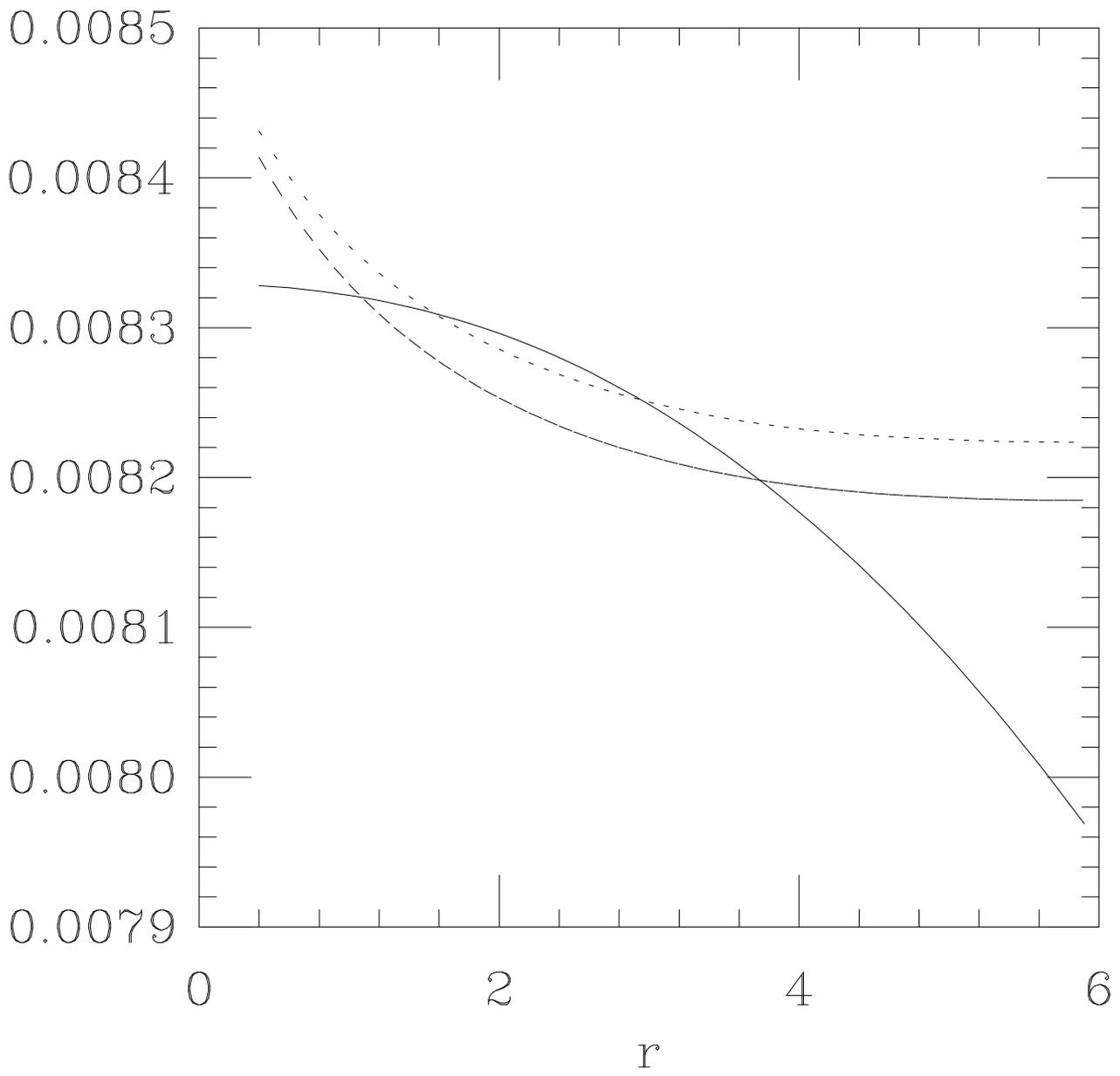
\end{document}